\documentclass[12pt,reprint,tightenlines,groupedaddress,superscriptaddress,floatfix,aps]{revtex4}
\usepackage{graphicx}
\usepackage{rotating}
\usepackage{array}
\usepackage{multirow}
\usepackage{epstopdf}
\usepackage{amsmath}
\usepackage{amsfonts}
\usepackage{setspace}
\usepackage{braket}
\usepackage{color}

\usepackage{hyperref}

\DeclareMathOperator{\Tr}{Tr}

\begin{document}
\title{Tensor hypercontraction for fully self-consistent imaginary-time GF2 and GWSOX methods: 
theory, implementation, and role of the Green's function second-order exchange for intermolecular interactions
} 

\author{Pavel Pokhilko}
\affiliation{Department  of  Chemistry,  University  of  Michigan,  Ann  Arbor,  Michigan  48109,  USA}
\author{Chia-Nan Yeh}
\affiliation{Center for Computational Quantum Physics, Flatiron Institute, New York, New York 10010, USA}
\author{Miguel A. Morales}
\affiliation{Center for Computational Quantum Physics, Flatiron Institute, New York, New York 10010, USA}
\author{Dominika Zgid}
\affiliation{Department  of  Chemistry,  University  of  Michigan,  Ann  Arbor,  Michigan  48109,  USA}
\affiliation{Department of Physics, University of Michigan, Ann Arbor, Michigan 48109, USA }
%\email{}

\renewcommand{\baselinestretch}{1.0}
\begin{abstract}

We apply tensor hypercontraction (THC) to reduce the computational scaling of expensive fully self-consistent Green's function methods. We present an efficient MPI-parallel algorithm and its implementation for evaluating the correlated second-order exchange term (SOX). This approach enabled us to conduct the largest fully self-consistent calculations with over 1100 atomic orbitals (AOs), with negligible errors attributed to THC fitting.
 
Utilizing our THC implementation for scGW, scGF2, and scGWSOX (GW plus the SOX term iterated to achieve full Green's function self-consistency), we evaluated energies of intermolecular interactions. This approach allowed us to circumvent issues related to reference dependence and ambiguity in energy evaluation, which are common challenges in non-self-consistent calculations.

We demonstrate that scGW exhibits a slight overbinding tendency for large systems, contrary to the underbinding observed with non-self-consistent RPA. Conversely, scGWSOX exhibits a slight underbinding tendency for such systems.
This behavior is both physical and systematic and is caused by exclusion-principle violating diagrams or corresponding corrections. 
Our analysis elucidates the role played by these different diagrams, which is crucial for the construction of rigorous, accurate, and systematic methods.
Finally, we explicitly show that all perturbative fully self-consistent Green's function methods are size-extensive. 
\end{abstract}
\maketitle

\section{Introduction}
Among the numerous variants of many-body perturbation theory, 
Green's function methods deserve a special consideration. 
On one hand, unlike wave functions, a one-particle Green's function is a compact object, 
requiring only two orbital indices and one time (or frequency) index for storage, 
making it an attractive platform for developing approximations for both molecules and extended systems. 
On the other hand, many physical laws can be rigorously formulated in terms of Green's functions, 
leading to a hierarchy of \emph{conserving} approximations fulfilling important physical properties\cite{Baym61,Baym62}. 
These powerful and systematically improvable theories have led to rigorous descriptions and accurate predictions 
of ionization and electron attachment, optical absorption, conductivity, magnetic properties, 
and finite-temperature phenomena\cite{Mahan00,Negele:Orland:book:2018,Martin:Interacting_electrons:2016}.

Traditionally, in Green's function approaches, certain diagrammatic terms are either truncated or simplified due to the expense of the the self-energy evaluation. 
For example, a full Green's function self-consistency is often replaced by various 
partial self-consistent schemes or even dropped entirely, such as is the case in $G_0 W_0$. 
While such non-self-consistent schemes are computationally cheaper, they often inherit a dependence on the used reference, 
which can be quite severe. 
This is especially known well for band gaps, which depend on a choice of selected orbitals 
(which are often taken from Kohn--Sham density functional approximations).  
Energy differences are even more sensitive to the reference dependence. 
For example, the dependence of broken-symmetry G$_0$W$_0$ on a DFT reference is so strong 
that it makes a non-self-consistent Green's function $G_0 W_0$ unusable for the evaluation of magnetic properties\cite{Chibotaru:BS-G0W0:2020}. 
Similarly, the random-phase approximation (RPA), which uses non-self-consistent Green's function, was employed for evaluating intermolecular interactions\cite{Scheffler:RPA:interactions:2011,Angyan:RPA:intermol:2010,Heselmann:RPA:intermol:2017} demonstrating
a strong dependence of intermolecular interaction energies on the reference orbitals.
Even an introduction of static and dynamic vertex insertions 
into the self-energy preserves a significant reference dependence\cite{Modrzejewski:RPA:intermol:2020,Forster:SOSSX:2022}. 
The dependence on orbital choice is not surprising because RPA and related methods are derivable from coupled-cluster doubles (CCD)\cite{Scuseria:RPA:CC:2008,Scuseria:ppRPA:CC:2013}.  
The dependence of CCD on the reference orbitals has been one of the motivations for introduction of singles excitations (coupled-cluster singles and doubles, CCSD) as well as Brueckner and orbital-optimized CCD variants\cite{BartlettShavitt:CC}. 
In addition to the reference dependence, the non-self-consistent methods are non-conserving~\cite{Baym61,Baym62}. In such  non-conserving Green's function methods
the evaluation of energy is not unique and depends on the way the energy is evaluated, leading to additional sources of errors and ambiguities.

There exist another way of reducing the computational cost of Green's function methods without diagrammatically simplifying the expressions for self-energy or Green's function. In these methods, a factorization and/or compression of tensors is used to reduce the computational cost without necessarily sacrificing the accuracy.
Factorization and compression techniques are very common within cost reduced wave-function approaches due their high computational expense and high accuracy requirements. 
Various wave-function compression schemes have been developed based on orbital transformations, 
which reduced the size of amplitude tensors, making treatments of larger molecules possible\cite{Bender:NO:67,Peyerimhoff:NO:74,Peyerimhoff:NO:77, Feller:NOvsHFinCI:92,CDS:98:Rev,Ruedenberg:NOvsHFinCI:02,Sherrill:NOsCI:03,Matsika:FNO:11,Matsika:FNO:13,BARTLETT:FNOvsOVOS:89,Bartlett:FNO:05,Bartlett:FNO:08,Pavel:OSFNO:2019}.
Reduced numerical representations also sped up contractions\cite{Martinez:GPU1:2008,Martinez:GPU2:2008,Martinez:GPU3:2008,Hammond2011:CCD,DePrince:RI-CCSD:GPU:2014,chaper:Eriksen:openacc:2017,Eriksen:openacc:2017,Rys:mp:2012,BRUSH:2015,Vogt:GPU:2008,Cederbaum:CD-MP2:sp:2011,Pavel:SP:2018,Crawford:SP:RT-CC:2022,Ma:SP:DMRG:2022,Wang:SP:CCSDpT:2020,Wang:SP:CCSDpT:2021}.
Exploitation of locality has reduced the computational cost further 
leading to even more efficient hybrid approaches\cite{Neese:locCEPA:2009,Kallay:locCCSDpT:2018,Neese:LocalCC:13,Neese:LocalPNOCCSD:2009,Neese:LocalT:2013,NeeseValeev:SparseMaps:2015,NeeseValeev:SparseMaps:2016,NeeseValeev:DLPNO-CCSD:2017,Neese:DLPNO:EOMIP:2018}.  
Resolution-of-identity (density fitting)\cite{Whitten:integrals:73,Dunlap:DF:1979,Eichkorn:RIbasis:95-0} 
and Cholesky\cite{Beebe:Cholesky:77,Koch:Cholesky:2003,Aquilante:Cholesky:2007,Koch:CholMethodSpec:2008,Aquilante:Cholesky:2009,Aquilante:Cholesky2:2009} 
decompositions of two-electron integrals are the most common decompositions  
implemented in most quantum chemical packages within a myriad of methods and properties\cite{Eichkorn:RIbasis:95-0,Weigend:RI_vs_Cholesky:09,Weigend:RIbasis:02,Hattig:RICC2:2000,Kohn:RICCDgrad:03,Hattig:RICCD:TP:02,Aquilante:CDMP2:07,Hattig:RiCC2:2003,Sherrill:RIGradCCSD_t:2017,Werner:RIMRMP2grad:2013,Lindh:RISACASgrad:2015,Sherrill:RIGradCCSD:2016,Krylov:RICholesky,Nanda:2PA:14,CD_gradient}. 
These decompositions reduce the size of integrals and reduce the scaling of \emph{some} terms, 
but often do not reduce the scaling of the overall method. 
To mitigate this problem, more sophisticated tensor decompositions have been introduced, 
such as the pseudospectral approach\cite{Friesner:PS:1985,Friesner:PS:1986,Martinez:PS:FCI:1992,Martinez:PS:MP3:1994}, 
the tensor hypercontraction (THC)\cite{Martinez:THC-MP2:2012,Martinez:LS-THC:2012,Martinez:LS-THC:3:2012,Martinez:THC-CC2:2013,Martinez:THC-CCSD:2014,Martinez:THC-MP2:2015,LinLin:ISDF:2017,Matthews:THC:2020,Lee:THC:2020,Yang:ISDF:THC:2023}, 
and the canonical polyadic decomposition\cite{Carroll:CP:1970,Hackbush:CP:MP2:2011,Auer:CP:CCSD:2013,Gruneis:CP:CC:2017,Valeev:CP:2021,Valeev:CP:2023}. 
Since spatial grids have led to relatively straightforward THC factorizations, 
they have received the most attention leading to the development of several variants of THC and 
numerous applications to HF, DFT, and correlated wave-function methods reducing their overall computational scaling\cite{LinLin:ISDF:2017,Martinez:THC-MP2:2012,Martinez:LS-THC:3:2012,Martinez:THC-CC2:2013,Martinez:THC-CCSD:2014,Martinez:THC-MP2:2015,Lee:THC:2020}.

In many Green's function's methods, a sacrifice of theoretical rigor is completely unnecessary for 
making these approaches more computationally affordable. 
Efficient time and frequency grids\cite{Kananenka:grids:2016,Kananenka16,Iskakov_Chebychev_2018,dong2020legendrespectral,Yoshimi:IR:2017} 
reduced the grid prefactor by orders of magnitude, 
enabling pioneering fully self-consistent Green's function calculations.
Efficient schemes using natural orbitals and localized orbitals\cite{Rusakov14,Rusakov:SEET:2019,Tran_generalized_seet}  
also has been utilized to reduce computational cost. 
RI decomposition deployed within scGW\cite{Iskakov20,Yeh:GPU:GW:2022,Yeh:X2C:GW:2022} 
led to a quarticly scaling fully self-consistent approach, 
the accuracy of which for magnetic interactions\cite{Pokhilko:local_correlators:2021,Pokhilko:BS-GW:solids:2022,Pokhilko:Neel_T:2022}, electron attachment, and ionization potentials\cite{Iskakov20,Yeh:GPU:GW:2022,Wen:2023,Abraham:X2C-GW:2024,Harsha:X2C-GW:2024} in molecules and solids 
is comparable to high-level wave-function methods.  
Recently deployed THC decomposition for scGW and RPA has reduced the scaling even further to a cubic one~\cite{Yeh:THC-GW:2024,Yeh:THC-RPA:2023}.  
Since RI cannot reduce the scaling of exchange terms, 
Green's function methods with correlated exchange, such as self-consistent Green's function second order (scGF2), have remained equally computationally expensive as without RI. 
Due to this issue, scGF2 has been applied only to relatively small molecules and solids\cite{Phillips14,Rusakov16,Welden16,GF2_Sergei19}.
The performance of this theory for larger systems remained unknown. 

In this paper, we introduce an MPI-parallel quarticly scaling THC algorithm for evaluating
the second-order exchange part (SOX) of self-energy that is also fully self-consistently evaluated. 
We show that systematically improvable THC decomposition leads to negligible errors in energies. 
This development has allowed us to make an assessment of THC-scGW, THC-scGF2, THC-scGWSOX 
(GW diagrams plus SOX diagram) methods and apply them to evaluate  intermolecular interactions. 
We also focus our discussion on interesting comparisons and connections with the wave-function theories. 
In particular, we confirm that scGF2 diverges quicker than MP2 for intermolecular interactions, while this does not happen with scGW and scGWSOX. The latter approaches CCSD for dispersion-dominated interactions, which is also consistent from a diagrammatic analysis. 
Finally, we show the size-extensivity of the Green's function methods, which is crucious for evaluation of total energies and total energy differences.

\section{Theory}
\subsection{Definitions}
The one-particle Green's function\cite{Mahan00,Negele:Orland:book:2018,Martin:Interacting_electrons:2016} is defined as
\begin{gather}
G_{pq} (\tau) = -\frac{1}{Z} \Tr \left[e^{-(\beta-\tau)(\hat{H}-\mu \hat{N})} p e^{-\tau(\hat{H}-\mu \hat{N})} q^\dagger  \right] \protect\label{eq:G_def}, \\
Z = \Tr \left[ e^{-\beta(\hat{H}-\mu \hat{N})} \right], 
\end{gather}
where $G$ is the one-particle Green's function, $Z$ is the grand-canonical partition function, 
$\hat{H}$ is the Hamiltonian operator, $\beta$ is the inverse temperature (in a.u.$^{-1}$), 
$\tau$ is the imaginary time, $\mu$ is the chemical potential, $\hat{N}$ is the particle-number operator, 
$p$ and $q^\dagger$ are the second-quantized creation and annihilation operators for the spin-orbitals p and q, respectively. 
The trace $\Tr$ in the expressions above is evaluated in the Fock space of all possible configurations. 
The knowledge of the exact one-particle Green's function provides access to many properties, 
such as all one-particle properties, total energies, ionization potentials and electron affinities. 

The Dyson equation is defined as
\begin{gather}
G^{-1}(i\omega_n) = G^{-1}_0(i\omega_n) - \Sigma[G](i\omega_n), \protect\label{eq:Dyson}\\
{G}^{-1}_0(i\omega_n) = i\omega_n +\mu \hat{N} - \hat{H}_0,
\end{gather}
where $G_0$ is the one-particle Green's function of independent electrons called the zeroth-order Green's function, 
$\omega_n$ is the Matsubara frequency,
$\Sigma[G]$ is the self-energy (with static and dynamic parts) that depends on the Green's function as a functional. 
The operator $\hat{H}_0$ is a one-body Hamiltonian operator for a system under study. In our case, $\hat{H}_0$ is the Hamiltonian of independent electrons. 
In non-orthogonal orbitals, such as atomic orbitals (AO), the Dyson equation has the same form 
and $G_0$ is written as
\begin{gather}
{\mathbf{G}}^{-1}_0(i\omega_n) = (i\omega_n+\mu) \mathbf{S} - \mathbf{H}_0,
\end{gather}
where we used the bold font for a representation of a matrix representation of operators in a specific spin-orbital basis set. 
The matrix $\mathbf{S}$ is the overlap matrix. 
The Dyson equation is not only satisfied for the exact Green's function and self-energy, 
it is also useful for approximations. 
We work with the fully self-consistent Green's function methods 
that approximate the self-energy functional as defined by Luttinger and Ward~\cite{Luttinger60}. 
Such approximations are \emph{conserving} and as shown by Kadanoff and Baym~\cite{Baym61,Baym62} lead to conservation of momenta, angular momenta, and energy; thermodynamic consistency; current continuity; and gauge invariance of the Luttinger--Ward functional.  
Such methods, in contrast to wave function methods, do not specify configurations explicitly in the Green's function definition~\ref{eq:G_def} 
and find the Green's function iteratively based on the Luttinger-Ward self-energy functional. 
The details of practical iterative algorithms used to reach self-consistency can be found in Refs~\cite{Pokhilko:algs:2022,Yeh:GPU:GW:2022, Iskakov20}. 
Once the self-consistency is achieved, the energy is evaluated using the Galitskii--Migdal formula~\cite{note:GM}
\begin{gather}
\gamma_{pq} = -G_{qp}(\tau=0^-), \\
E_{\infty} = \sum_{pq} \left(h_{pq}+\frac{1}{2}\Sigma^{HF}_{pq}[G]\right) \gamma_{pq}, \protect\label{eq:E1b}\\
E_{dyn} = \frac{1}{\beta}\sum_{\omega_n} \sum_{pq} G_{qp}(\omega_n) \Sigma_{pq}^{dyn}[G](\omega_n), \protect\label{eq:E2b}
\end{gather}
where $\gamma$ is the one-particle density matrix, 
$E_{\infty}$ and $E_{dyn}$ are the HF energy evaluated from correlated one-electron density matrix and dynamical (two-body) energy parts, respectively. 
The total energy is a sum of  one-body and two-body energies, $E_{tot}=E_{\infty}+E_{dyn}$.
The fully self-consistent Green's function methods, as analyzed by Baym and Kadanoff~\cite{Baym61,Baym62}, are conserving that means that different ways of evaluating total energy yield the same results. This is especially important when analyzing energy differences which are uniquely defined in conserved methods.

\begin{figure}[!h]
  \includegraphics[width=8cm]{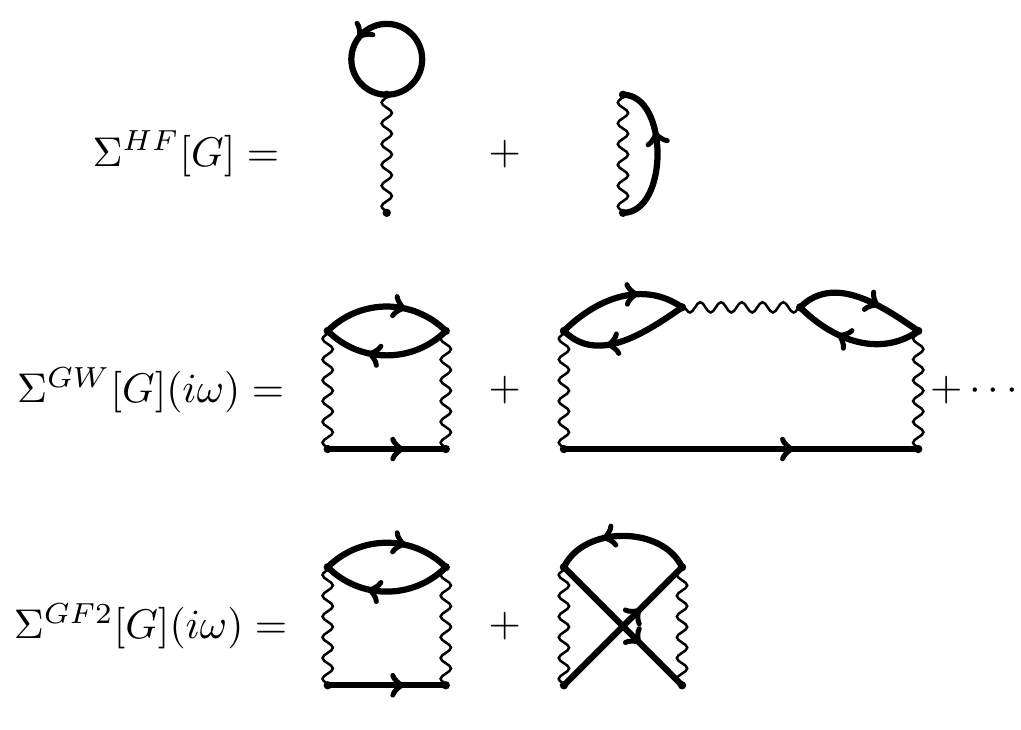}
\centering
\caption{Skeleton self-energy series: static (HF) self-energy diagrams (top), dynamical self-energy diagrams of GW (middle) and GF2 (bottom). The propagator lines are bold, indicating a \emph{full} renormalized one-particle Green's function $G$ iterated to the full self-consistency. The curvy lines indicate a \emph{bare} interaction $v$ (non-antisymmetrized two-electron integrals). 
         \protect\label{fig:diag}
}
\end{figure}

In this work, we construct all the analyzed self-energy functionals perturbatively. Consistently with our previous publications, we use the \emph{bare} perturbation theory with the \emph{bold} propagators. The number of two-electron integrals (\emph{bare} interaction lines) in the self-energy functional expression determines the perturbative order of such diagrams in terms of \emph{bare} interactions. 
The bold (thick) propagator lines denote full self-consistency by the Green's functions and a full self-consistent solution of the Dyson equation with the update of the chemical potential. 
Figure~\ref{fig:diag} shows diagrammatic relationships between different methods. 
The Hartree--Fock theory contains only Coulomb and exchange terms. The Hartree--Fock self-energy is static (frequency-independent). 
Correlated fully self-consistent methods also have a dynamical part of the self-energy functional responsible for correlation. 
scGF2\cite{Snijders:GF2:1990,Dahlen05,Phillips14,Rusakov16,Welden16,Iskakov19} is the self-consistent second-order  perturbation theory, where the self-energy is constructed up to the second order in \emph{bare} interactions
containing both correlated direct and exchange terms shown in Fig.~\ref{fig:diag}. 
scGW\cite{Hedin65,G0W0_Pickett84,G0W0_Hybertsen86,GW_Aryasetiawan98,Stan06,Koval14,scGW_Andrey09,GW100,Holm98,QPGW_Schilfgaarde,Kutepov17,Iskakov20,Yeh:X2C:GW:2022,Yeh:GPU:GW:2022} is the first-order \emph{screened} perturbation theory, where the self-energy is constructed from a single screened interaction $W$, which includes an infinite series of bubble diagrams, leading to self-energy diagrams shown in Fig.~\ref{fig:diag}. 
The direct second-order term is included in both scGF2 and scGW. 
scGW lacks correlated exchange terms; therefore, it violates Pauli exclusion principle  
and breaks permutational properties of two-particle density matrices\cite{note:crossing_sym}, explaining why its total energies are often below full configuration interaction (FCI)\cite{note:ed} within a given basis set.

The addition of the scGF2 exchange diagram to scGW self-energy expression has two major advantages
\begin{itemize}
    \item it completes the second order of self-energy in \emph{bare} interactions and maintains Pauli exclusion principle up to the second order and 
    \item it approximates an insertion of a 3-point vertex function into the self-energy expression.
\end{itemize}
The resulting self-energy functional, which we call scGWSOX in this manuscript, contains all the terms from scGW and scGF2.

\subsection{Tensor hypercontraction}
As originally proposed in Ref.~\citenum{Martinez:THC-MP2:2012}, 
THC is described as a factorized representation of electron repulsion integrals (ERIs), illustrated by the following equation:
\begin{gather}
(ij|kl) \approx \sum_{PQ}^{N} (X_{iP} )^* X_{jP} U_{PQ} (X_{kQ})^* X_{lQ}, 
\protect\label{eq:THC_ERI}
\end{gather}
where $\textbf{X}$ denotes the collocation matrix, and uppercase letters $(P, Q)$ refer to the indices within the THC interpolation vectors. 
Many different schemes have been proposed to achieve such a decomposition~\cite{Martinez:THC-MP2:2012,Martinez:LS-THC:2012,Parrish:THC_DVR:2013,Martinez:THC-MP2:2015,Lu:ISDF:2015,Lu:ISDF_Bloch:2016}. 

In this work, we adapt the least-squares THC (LS-THC) method~\cite{Martinez:LS-THC:2012} within the atomic orbitals (AOs) framework. 
Here, the collocation matrix $X_{iP}$ is defined as the AO basis function $\phi_{i}(\textbf{r}_{P})$ evaluated at a selected set of real-space points $\{\textbf{r}_{P}\}$. 
For practical implementation, these grid points are often pre-optimized for each atomic species based on the given AO basis. 
In the current study, we construct $\{\textbf{r}_{P}\}$ by pruning an initial Becke-style molecular grid~\cite{Becke:mo_grid:1988} using the Cholesky-based methods~\cite{Matthews:THC:2020} for each molecule on the fly. 

The remaining matrix $\textbf{U}$ is subsequently evaluated by least-squares fitting to the ERIs: 
\begin{gather}
L^{C}_{ij} = \sum_{P}X_{iP}^{*}X_{jP}A_{P}^{C}
\end{gather}
where $L^{C}_{ij}$ is the three-index tensor obtained through the density fitting of ERIs\cite{RIpaper1}: 
\begin{gather}
(ij|kl) \approx \sum_{C}L^{C*}_{ji}L^{C}_{kl}. 
\end{gather}
Once $A^{C}_{P}$ is obtained, the matrix $\textbf{U}$ is evaluated as 
\begin{gather}
U_{PQ} = \sum_{C}A^{C*}_{P} A^{C}_{Q}. 
\end{gather}

\subsection{THC-scGF2: algorithm and implementation details}
\begin{figure}[!h]
  \includegraphics[width=8cm]{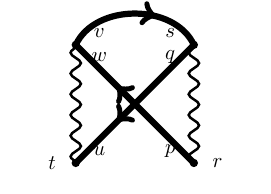}
\caption{A labeled scGF2 exchange diagram (SOX) with bold Green's function lines.
         \protect\label{fig:GF2_ex}
}
\end{figure}
The treatment of the GF2 direct term withing the THC formalism is nearly identical to the treatment of the GW diagrams from Ref.\cite{Yeh:THC-GW:2024}, which we fully reused in our code. 

The SOX diagram from scGF2, depicted in Fig.~\ref{fig:GF2_ex}, which contains bold Green's function lines, deciphers algebraically as
\begin{gather}
\Sigma_{tr}^{SOX}(\tau_{tr}) = -\sum_{pqsuvw}(pr|qs)(tu|vw) G_{wp}(\tau_{wp}) G_{uq} (\tau_{up}) G_{sv}(\tau_{sv}),
\end{gather}
where $(pr|qs)$ are the ERIs in the chemical notation. 
Since the \emph{bare} interaction is instantaneous (ERIs do not depend on time), the following time relationships are fulfilled
\begin{gather*}
\tau_t = \tau_u = \tau_w = \tau_v \\
\tau_p = \tau_r = \tau_q = \tau_s 
\end{gather*}
We can now simplify the time indices:
\begin{gather*}
\tau_{wp} = \tau_{uq} = \tau \\
\tau_{sv} = -\tau.
\end{gather*}
This results in the well known expression 
\begin{gather}
\Sigma_{tr}^{SOX}(\tau) = -\sum_{pqsuvw}(pr|qs)(tu|vw) G_{wp}(\tau) G_{uq} (\tau) G_{sv}(-\tau).
\end{gather}
When this term is evaluated with the help of several intermediate quantities (corresponding to only one contraction with $G$ at a time as in standard ERI transformation algorithms), its scales as $O(n_{\tau} n^5_{AO})$, where $n_{\tau}$ is the number of time grid points, $n_{AO}$ is the number of atomic orbitals. The scaling of this term cannot be reduced by density fitting, but can be reduced with THC, which we explore in this work.
We use the THC algebraic expression shown in Eq.~\ref{eq:THC_ERI}. 
While there could be multiple ways how THC-SOX can be implemented, 
one should keep in mind that the number of interpolating vectors $N$ can be quite large. 
Therefore, if a desired computational scheme aims to minimize memory usage, the intermediate quantities evaluated should have many indices in the original basis (we use AO) and only a few in the basis of interpolated vectors.
In our scheme, lower case labels (p,q,r,s,t,u,v,w) are denoting atomic orbital indices while the upper case labels (P,Q) are  used for the interpolated vectors.
We implemented the scheme shown below
\begin{gather}
U_{Pqs} = \sum_Q U_{PQ}(X_{qQ})^* X_{sQ}\qquad O(N^2 n^2_{AO}), \protect\label{eq:U_contraction} \\
G_{Rq} (\tau) = \sum_{uq} X_{uR} G_{uq}(\tau) \qquad O(N n_\tau n^2_{AO}),  \\
G_{wP} (\tau) = \sum_{uq} G_{wp}(\tau) (X_{pP})^* \qquad O(N n_\tau n^2_{AO}), \\
A_{Pqv} = \sum_s U_{Pqs} G_{sv} (-\tau) \qquad O(N n_\tau n^3_{AO}), \\
B_{PRv} = \sum_q A_{Pqv} G_{Rq} (\tau) \qquad O(N^2 n_\tau n^2_{AO}), \\
C_{RvP} = \sum_w U_{Rvw} G_{wP} (\tau)  \qquad O(N^2 n_\tau n^2_{AO}), \\
D_{PR} = \sum_v C_{RvP} B_{PRv} \qquad O(N^2 n_\tau n_{AO}), \\
\Sigma_{RP}(\tau) =  -D_{PR} \qquad O(N^2 n_\tau), \\
\Sigma_{tr}(\tau) = \sum_{PR}  (X_{tR})^* \Sigma_{RP}(\tau) X_{rP} \qquad O(N^2 n_\tau n_{AO}),
\end{gather}
where only the Green's function and the self-energy are stored at all time points. 
Since the SOX self-energy expression is linear in its time dependence, 
we store the 3-index intermediates only at a running time point to save memory. 
For tensor representations, we used \textsc{NDA} library\cite{nda}. 
We implemented the first contraction (Eq.~\ref{eq:U_contraction}) using \textsc{SLATE} distributed linear algebra library\cite{SLATE:2019}, 
which uses a parallel algorithm for distributed matrix multiplications through a sequence of outer products.  
We implemented the subsequent contractions building $A$, $B$, $C$, and $D$ using \textsc{NDA} BLAS linkings. 
All of the contractions are MPI-parallel over at least one auxiliary index. 

In our experience, the initial contraction operation in Eq.~\ref{eq:U_contraction}, which entails a matrix multiplication by the X-X outer product, proves to be the most costly.
However, it does not contain any Green's functions, so it does not need to be recomputed during the self-consistency when Green's function is updated. 
Consequently, we evaluate it only once and it is re-used during all the self-consistent iterations. Our implementation will be available as an open-source package in the future together with THC-GW and THC-RPA implementations in Refs.~\cite{Yeh:THC-GW:2024,Yeh:THC-RPA:2023}.

Since the number of interpolating points $N$ can be substantial, 
it can lead to an overall cost increase. 
Therefore, it is important to pay attention to how an algorithm scales not only 
with respect to overall number of basis functions $n_{AO}$, but also to $N$. 
All the operations in our exchange algorithm are at most quadratic with respect to $N$. 
This is an important distinction from an earlier THC-MP2 exchange implementation\cite{Lee:THC:2020}, 
which has a cubic scaling with respect to the number of interpolating points $N$. 
Other THC-MP2 exchange implementations\cite{Martinez:THC-MP2:2012,Martinez:THC-MP2:2015,Matthews:THC:2020} 
scale as $O(n_l n_o n_v N^2)$, where $n_l$ is the number of grid points for the Laplace transform, 
$n_o$ is the number of occupied orbitals, 
$n_v$ is the number of virtual orbitals. 
A close inspection of their algorithmic procedure reveals a similarity with our THC-GF2 implementation in the operations and intermediates used. 
One of the substantial differences is due the lack of  separation of  the occupied and virtual orbitals in our implementation. 
This is because in our finite temperature Green's function formalism  a clear separation between occupied and virtual orbitals cannot be executed since at different temperatures different orbitals can become occupied or unoccupied and all are treated at the same footing. 
In addition, the size of the imaginary-time grid used for Green's function calculations is usually larger 
(by one order of magnitude or more) than the grid used for the Laplace transform in MP2. 
This is because grids used in non-self-consistent approaches are optimized for a specific integral or even a specific system. 
The grids that we use are general and can support any time-dependent correlator, such as the Green's function. Because of this, such grids can be successfully used to carry out full self-consistency.
Although our current work focuses on the SOX term with the \emph{bare} interactions, contractions of the same type will also appear in the theories with \emph{screened} interaction insertion, such as second-order screened exchange (with one \emph{screened} interaction line) and G3W2 (with two \emph{screened} interaction lines). We will pursue such \emph{screened} theories in the near future.

\section{Numerical results}
\subsection{Computational details}
The geometries of the systems are taken from the original S22 benchmark\cite{Hobza:bench:06} (version A), 
which also have been used in the subsequent revisions in Refs\cite{Sherrill:S22:2010,Sherrill:SCS-CCSD:NCI:2014}.  
The ISDF was performed with guess seeding from Becke quadrature grids\cite{Becke1988} 
similarly to Ref.\cite{Lee:THC:2020}. 
We used aug-cc-pVTZ basis set\cite{Dunning:ccpvxz:1989,Dunning:aug-ccpvxz:1992} in all calculations. 
All the integrals were prepared with PySCF program\cite{PYSCF}.
To perform the MP2 and OOMP2 calculations, we used the corresponding codes in PySCF\cite{OOMP2_code}. 
For the Green's function calculations, we used the resolution-of-identity decomposition\cite{RIpaper1} of two-electron integrals 
with even-tempered auxiliary functions using PySCF generator with $\beta=1.5$. 
We used the inverse temperature of $1000$~a.u.$^{-1}$ and the intermediate representation with 
$\Lambda=10^5$ from \emph{sparse-ir} package\cite{Sparse-ir:grid:2023}  
for all Green's function calculations, 
which we converged very tightly to resolve small energy differences reported in the paper. 
We used damping and frequency-dependent commutator DIIS algorithm from Ref\cite{Pokhilko:algs:2022} designed specifically for the fully self-consistent Green's function calculations.
We performed all-electron Green's function calculations since at the moment the frozen-core formalism is not developed for self-consistent Green's functions at a finite temperature. 
Unfortunately, the S22 reference benchmark papers\cite{Sherrill:S22:2010,Sherrill:SCS-CCSD:NCI:2014} do not indicate whether the core electrons were frozen. 
We checked a few small molecules from the S22 set.  
Only the frozen-core CCSD and CCSD(T) matched the previously reported results. 
To make a clear distinction between all-electron and frozen-core calculations, 
we write abbreviate the latter with ``fc'' prefix. 
A comparison between all-electron and frozen-core calculations for a few selected systems is shown in SI in Table S1, which shows that the all-electron and frozen-core estimates without counterpoise correction (CP)\cite{Boys:70} are substantially different. The counterpoise correction mostly removes this discrepancy making comparisons between all-electron and frozen-core calculations possible. This observation is consistent with the previous benchmark results\cite{Furche:RPA:S22:2012}. Unless noted explicitly, counterpoise correction is included in all the reported interaction energies.

\subsection{THC convergence of interaction energies}
\begin{table*} [tbh!]
  \caption{RI- and THC-HF interaction energies (kcal/mol) for ammonia dimer, with and without a counterpoise correction. 
           Different values of $\alpha$ are shown for THC.}
\protect\label{tbl:HF_NH3}
\begin{tabular}{l|cc}
\hline
\hline
ints          & w/o CP   & w CP \\
\hline
RI            & -1.43  & -1.40   \\
\hline
$\alpha = 20$ & -503   & -502  \\
$\alpha = 21$ & -47.05 & -47.05  \\
$\alpha = 22$ & -1.48  & -1.42  \\
$\alpha = 23$ & -1.41  & -1.39  \\
$\alpha = 24$ & -1.39  & -1.41  \\
\hline
\hline
\end{tabular}
\end{table*}
\begin{table*} [tbh!]
  \caption{Interaction energies (kcal/mol) for small dimers, with and without a counterpoise correction. 
           Different values of $\alpha$ are shown for THC.}
\protect\label{tbl:GW_NH3}
\begin{tabular}{l|cc|cc|cc|cc}
\hline
\hline
(NH$_3$)$_2$  & \multicolumn{2}{c|}{scGW} & \multicolumn{2}{c|}{scGWSOX} & \multicolumn{2}{c|}{scGF2dir} & \multicolumn{2}{c}{scGF2} \\ 
              & w/o CP & w CP         & w/o CP & w CP    & w/o CP & w CP    & w/o CP & w CP  \\
\hline                                                                      
$\alpha = 5$  & -3.58 & -3.30         & -3.18  & -2.99   & -5.17  & -5.16   & -4.42  & -4.24 \\
$\alpha = 6$  & -3.66 & -3.07         & -3.17  & -2.85   & -5.27  & -4.63   & -4.43  & -4.07 \\
$\alpha = 7$  & -3.54 & -2.95         & -3.10  & -2.80   & -5.02  & -4.36   & -4.26  & -3.91 \\
$\alpha = 8$  & -3.58 & -2.86         & -3.13  & -2.73   & -5.01  & -4.24   & -4.26  & -3.81 \\
$\alpha = 9$  & -3.58 & -2.87         & -3.12  & -2.73   & -5.03  & -4.26   & -4.26  & -3.82 \\
$\alpha = 10$ & -3.57 & -2.87         & -3.12  & -2.74   & -5.01  & -4.25   & -4.25  & -3.82 \\
\hline
(H$_2$O)$_2$  & \multicolumn{2}{c|}{scGW} & \multicolumn{2}{c|}{scGWSOX} & \multicolumn{2}{c|}{scGF2dir} & \multicolumn{2}{c}{scGF2} \\ 
              & w/o CP & w CP         & w/o CP & w CP    & w/o CP & w CP    & w/o CP & w CP  \\
\hline                                                                      
$\alpha = 5$  & -6.73 & -5.18         & -6.06  & -5.23   & -8.29  & -6.65   & -7.18  & -6.26 \\
$\alpha = 6$  & -6.27 & -4.85         & -5.76  & -4.98   & -7.63  & -6.14   & -6.72  & -5.86 \\
$\alpha = 7$  & -6.20 & -4.73         & -5.70  & -4.91   & -7.41  & -5.81   & -6.55  & -5.65 \\
$\alpha = 8$  & -6.28 & -4.76         & -5.76  & -4.93   & -7.44  & -5.82   & -6.57  & -5.66 \\
$\alpha = 9$  & -6.27 & -4.73         & -5.75  & -4.92   & -7.41  & -5.74   & -6.55  & -5.60 \\
$\alpha = 10$ & -6.26 & -4.73         & -5.75  & -4.92   & -7.39  & -5.73   & -6.55  & -5.60 \\
\hline
(CH$_4$)$_2$  & \multicolumn{2}{c|}{scGW} & \multicolumn{2}{c|}{scGWSOX} & \multicolumn{2}{c|}{scGF2dir} & \multicolumn{2}{c}{scGF2} \\ 
              & w/o CP & w CP         & w/o CP & w CP    & w/o CP & w CP    & w/o CP & w CP  \\
\hline                                                                      
$\alpha = 5$  & -1.28 & -0.53         & -0.70  & -0.33   & -2.26  & -1.43   & -1.48  & -1.05 \\
$\alpha = 6$  & -1.21 & -0.54         & -0.64  & -0.31   & -2.16  & -1.45   & -1.40  & -1.02 \\
$\alpha = 7$  & -1.14 & -0.50         & -0.64  & -0.29   & -2.04  & -1.34   & -1.35  & -0.95 \\
$\alpha = 8$  & -1.12 & -0.49         & -0.60  & -0.28   & -2.01  & -1.34   & -1.30  & -0.94 \\
$\alpha = 9$  & -1.16 & -0.48         & -0.62  & -0.27   & -2.05  & -1.31   & -1.33  & -0.93 \\
$\alpha = 10$ & -1.15 & -0.48         & -0.63  & -0.27   & -2.03  & -1.31   & -1.33  & -0.93 \\
\hline
\hline
\end{tabular}
\end{table*}

In Table~\ref{tbl:HF_NH3}, we show the convergence with respect to  different values of $\alpha = \frac{N}{n_{AO}}$ for the THC-HF interactions energies for an
ammonia dimer. 
It is clear that only very high values of $\alpha$ can accurately recover energy differences. 
Such high values lead not only to computationally demanding calculations, 
but also to numerical failures of finding interpolation vectors due to their close linear dependence. 
Since the correlation energy is orders of magnitude smaller than the Hartree-Fock energy, a significantly more accurate approximation can be expected if THC is applied solely to the dynamical part of the self-energy, namely $\Sigma[G](i\omega_n)$, while the Fock matrix is evaluated without THC-decomposed integrals (RI in our case).
Within the wave-function approaches, similar results were observed with THC before\cite{Lee:THC:2020} as well as with mixed-precision approaches\cite{Martinez:mixed_prec:ints:2011,Pavel:SP:2018}.
In Table~\ref{tbl:GW_NH3}, we present interactions energies evaluated in such a manner with different self-consistent diagrammatic approaches. 
The convergence of interaction energies is achieved at much smaller values of $\alpha$. Calculations with and without counterpoise correction show a similar convergence behavior with respect to $\alpha$ indicating that the procedure is resilient with respect to the presence of functions on ghost atoms. 
In particular, $\alpha=10$ can be considered as fully converged. This value of $\alpha=10$ will be used in all other calculations performed by us.  

\subsection{Total energies}
\begin{table*} [tbh!]
  \caption{Total energies (in a.u.) of a few small selected dimers. All the calculations in the table used the same even-tempered auxiliary basis sets for a fair comparison between methods. No orbitals were frozen.
  }
\protect\label{tbl:tot_energies}
\begin{tabular}{l|ccc}
\hline
\hline
Method              & (NH$_3$)$_2$   & (H$_2$O)$_2$ & (CH$_4$)$_2$ \\
\hline
MP2           & -112.960328    & -152.697224  & -80.86716988 \\
OOMP2         & -112.964138    & -152.702513  & -80.86970025 \\
THC-scGF2     & -112.965596    & -152.701647  & -80.87265605 \\
THC-scGF2dir  & -113.256121    & -152.995941  & -81.15475658 \\
THC-scGW      & -113.088797    & -152.806993  & -81.02352119 \\
THC-scGWSOX  & -112.798849    & -152.513123  & -80.74176872 \\
CCSD          & -112.984118    & -152.706444  & -80.90895825 \\
\hline
\hline
\end{tabular}
\end{table*}
Total energies can reveal useful insights into physical meaning of diagrams of different types. 
Table~\ref{tbl:tot_energies} shows total energies of a few selected dimers evaluated with all-electron calculations. 
The orbital optimization procedure converges in a very few iterations, 
which is not surprising since the separation between the ground state and the excited states is large. 
The OOMP2 total energies are substantially lower than the MP2 total energies, 
indicating that the MP2 energies are sensitive to the choice of reference orbitals. 
Although the scGF2 and OOMP2 total energies are close for both dimers and monomers, they are not the same. 
While a comparison between wave-function perturbation theories 
and a quasiparticle qpGF2 has been done recently\cite{Tew:qpGF2:reg:2023},  
a careful comparison between OOMP2 and scGF2 has not been performed. We list a few key distinctions below. 
First, while the OOMP2 energy is stationary only with respect to static one-electron orbital rotations, 
the GF2 grand potential is stationary with respect to the full imaginary frequency dependent Green's function, 
which includes more variational degrees of freedom then just the static reference determinant transformations. 
Second, we use the OOMP2 formulation from Bozkaya\cite{Bozkaya:OOMP2:2011} dropping $T_1$ contributions 
(which are small but are not zero); thus OOMP2 does not have a complete perturbative second order. 
The second order included in scGF2 is complete.

If only the direct correlated GF2 term (GF2dir) is included into the self-energy functional, 
the total energy  decreases significantly  below CCSD. 
This happens due to a strong violation of fermionic permutational properties of the two-particle density matrix\cite{note:crossing_sym}. 
The full scGF2 (with both correlated direct and exchange terms) recovers the permutational property of the two-particle matrix and raises the total energy. 
The higher-order direct terms included in scGW partly suppress the effect of the direct second-order term. 
This again can be seen from the structure of the two-particle density matrix. 
scGW renormalizes the \emph{bare} Coulomb interaction by an effective dielectric constant 
replacing of the \emph{bare} interaction by the \emph{screened} interaction. 
Therefore, the scGW cumulant includes this renormalization, which increases the total energy from the scGF2dir to scGW. 
Inclusion of the SOX term in scGWSOX further increases the energy compared to scGW.
Since the renormalized interactions are not included in the second-order exchange term, it overcorrects the violation of permutational symmetry by scGW. 
Nonetheless, the information gained from scGWSOX is valuable since it sheds light on the impact of vertex corrections in a rigorous and fully self-consistent manner, eliminating large uncertainties stemming from a severe dependence on the reference orbitals and ambiguity in the energy evaluation.

\subsection{Interaction energies}
\subsubsection{A comparison with non-self-consistent methods}
\begin{figure}[!h]
  \includegraphics[width=11cm]{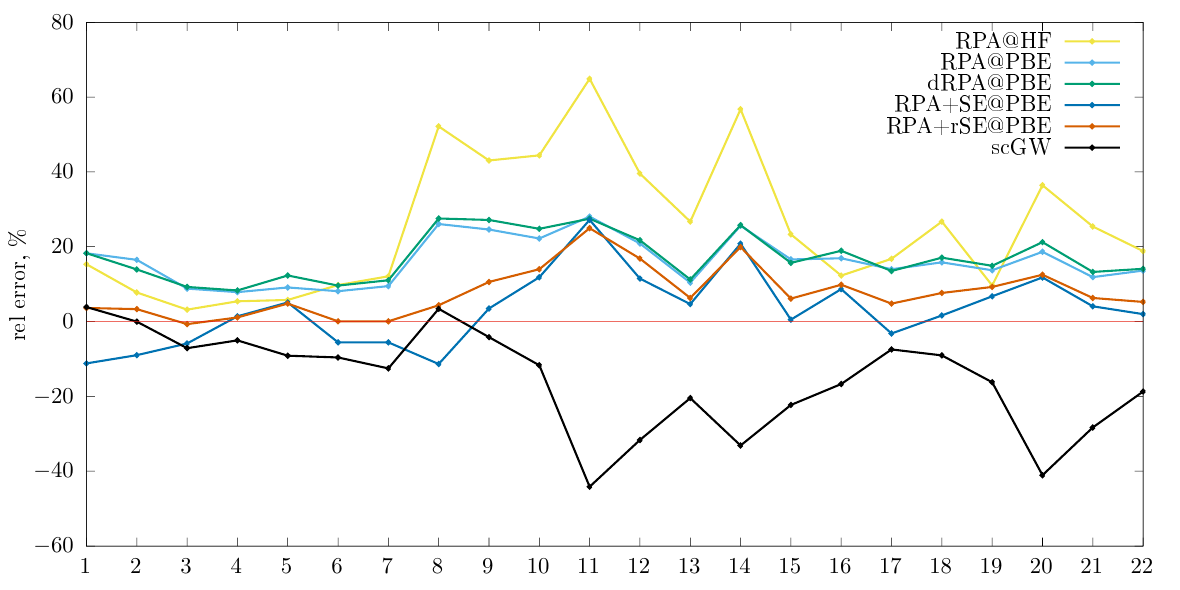}
\centering
\caption{A comparison between scGW and RPA-related methods. Only relative errors of interaction energies with respect to fcCCSD(T) references are shown. The negative values correspond to overbinding; the positive values correspond to underbinding. The RPA values from HF and PBE references (RPA@HF and RPA@PBE, respectively) are taken from Ref.\cite{Ren:RI-RPA:2012} and plotted relative to fcCCSD(T)/CBS from Ref.\cite{Sherrill:S22:2010}. The dRPA values from PBE reference at a custom augmented triple-zeta basis set are taken from Ref.\cite{VandeVondele:dRPA:2013} and plotted relative to fcCCSD(T)/aug-cc-pVTZ from Ref.\cite{Sherrill:S22:2010}, which delivers a comparable basis set quality. The relative errors of RPA+SE and RPA+rSE, are taken from Ref.\cite{Ren:RPA_SE:2013} and digitized using WebPlotDigitizer\cite{WebPlotDigitizer:2022}.
         \protect\label{fig:rpa_vs_gw}
}
\end{figure}

Size-consistency and size-extensivity are crucial properties for evaluation of energy differences since otherwise non-size-consistent errors can become dominant for large systems (and often are as shown in Ref.\cite{OlsenText}). We formally prove size-extensivity of the fully self-consistent methods in Appendix. 
In Figure~\ref{fig:rpa_vs_gw}, we compare the fully self-consistent THC-scGW with non-self-consistent RPA methods. For different methods, we plot \emph{relative} errors of counterpose-corrected interaction energies with respect to fcCCSD(T) references with comparable basis sets. 
We decided to plot the relative errors for different methods since the basis sets used in numerous RPA publications are different from each other and therefore the results of these studies cannot be compared against each other directly. 
The relative error comparison was used before in Ref.\cite{Ren:RPA_SE:2013} demonstrating that such a comparison  is essentially insensitive with respect to the basis set used, allowing us to make comparisons with our scGW results. The RPA interaction energies are very sensitive with respect to the reference orbitals used, which for some dispersion-dominated systems even changes the relative error by a factor of 2. Both RPA@HF and RPA@PBE underbind all systems from S22 dataset. The lowest-order perturbative correction for orbital relaxation with single excitations (+SE and +rSE, where the later is the semi-canonicalized variant of the former) increases the strengths of the interactions with occasional overestimation of interactions. 
Our fully self-consistent THC-scGW predicts much stronger intermolecular interactions and is completely reference-independent. The energy evaluation in scGW is also non-ambiguous since scGW is a conserving approximation. Thus, a complete removal of the reference dependence and ambiguities in the energy evaluation from RPA through full Green's function self-consistency completely changes the physical predictions. 
Additive perturbative corrections with single excitations (+SE and +rSE) introduced in RPA to capture orbital relaxation approximately lead to changes comparable by magnitude with the reference dependence of parent RPA. Such a behavior of these corrections also indicates to a strong dependence of RPA approaches on the selected reference. Although structurally scGW looks similar to RPA, scGW is theoretically superior to RPA since scGW maintains conservation laws and gauge invariance of the Luttinger--Ward functional making scGW independent on a selected reference. While +SE@PBE and +rSE@PBE corrections shift RPA estimates toward scGW, these additive corrections capture orbital relaxation only approximately without invoking exponential orbital rotations. Thus, such additive corrections cannot fully remedy the reference dependence problem of RPA.
While RPA results have smaller relative errors, one can suspect that this is a result of fortuitous cancellations. 
We fully explain the overbinding behavior of scGW in the next section.
\begin{figure}[!h]
  \includegraphics[width=11cm]{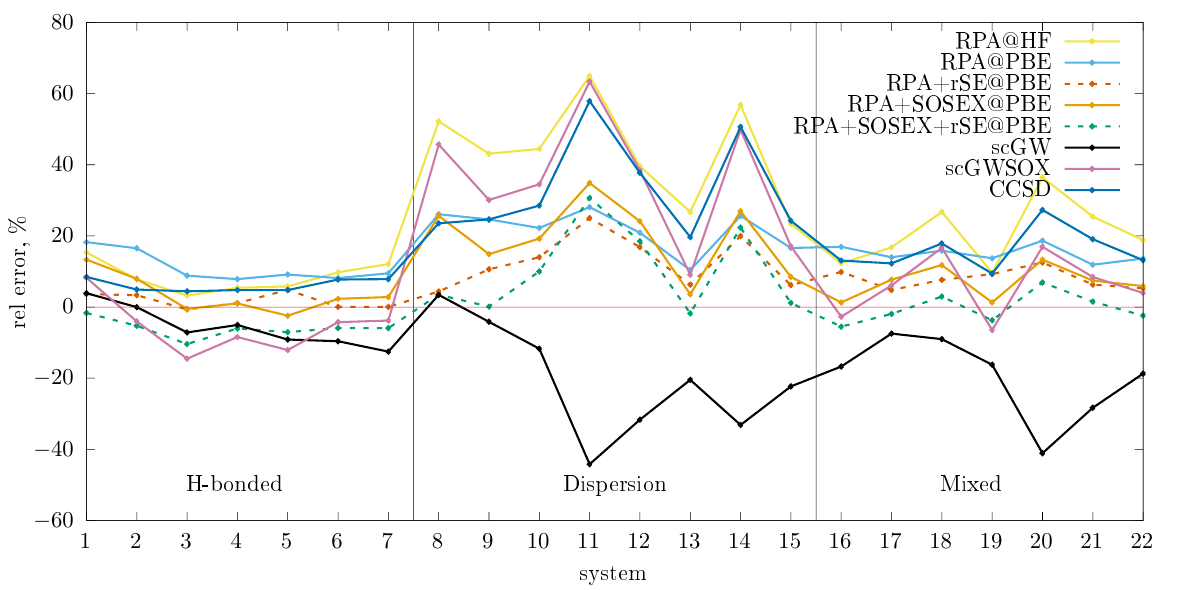}
\centering
\caption{A comparison between scGW and scGWSOX and non-self-consistent methods. Only relative errors of interaction energies with respect to fcCCSD(T) references are shown. The negative values correspond to overbinding; the positive values correspond to underbinding. In addition to RPA@HF and RPA@PBE from Ref.\cite{Ren:RI-RPA:2012}, RPA+rSE@PBE, 
RPA+SOSEX@PBE, RPA+SOSEX+rSE@PBE from Ref.\cite{Ren:RPA_SE:2013} are also shown relative to CCSD(T)/CBS. scGW, scGWSOX, and CCSD are shown relative to CCSD(T)/aug-cc-pVTZ.
         \protect\label{fig:sosex_vs_sox}
}
\end{figure}

We show additional comparisons in Figure~\ref{fig:sosex_vs_sox}. We see that addition of second-order screened exchange to RPA\cite{Ren:RPA_SE:2013} leads to smaller changes in interaction energies than the choice of the selected reference. Therefore, the accuracy of RPA+SOSEX is entirely dominanted by a strong reference dependence. While scGW systematically overbinds systems, scGWSOX systematically underbinds dispersion-dominated and mixed systems. Signed errors of CCSD are close to the ones of scGWSOX. We discuss this fact in detain in the next section.

\subsubsection{Analysis of fully self-consistent methods}
\begin{table*} [tbh!]
  \caption{S22 counterpoise-corrected interaction energies (kcal/mol) computed in the aug-cc-pVTZ basis set. The systems are listed preserving the enumeration from the original dataset. Number of AOs and interaction energies with self-consistent methods and references are shown. }
\protect\label{tbl:int_energies}
\begin{tabular}{ll|l|cccc}
\hline
\hline
\# & H-bonded                     &AOs & THC-scGW  & THC-scGWSOX  & fcCCSD$^a$  & fcCCSD(T)$^b$   \\
\hline                                     
1 & (NH$_3$)$_2$                  & 230  & -2.87  & -2.74  & -2.733  & -2.986      \\ 
2 & (H$_2$O)$_2$                  & 184  & -4.73  & -4.92  & -4.496  & -4.729      \\
3 & Formic acid dimer             & 368  & -19.02 & -20.33 & -16.970 & -17.755     \\
4 & Formamide dimer               & 414  & -16.06 & -16.58 & -14.559 & -15.291     \\
5 & Uracil dimer                  & 920  & -21.67 & -22.26 & -18.907 & -19.858     \\
6 & 2-pyridoxine 2-aminopyridine  & 897  & -17.78 & -16.91 & -14.969 & -16.222     \\
7 & Adenine-thymine WC            & 1127 & -17.94 & -16.55 & -14.691 & -15.943     \\
\hline                                     
  & Dispersion dominated          &AOs & THC-scGW & THC-scGWSOX & fcCCSD$^a$  & fcCCSD(T)$^b$    \\
\hline                                     
8 & (CH$_4$)$_2$                  & 276  & -0.48   & -0.27 & -0.380  & -0.497       \\
9 & Ethene dimer                  & 368  & -1.43   & -0.96 & -1.035  & -1.373       \\
10& Benzene-CH$_4$                & 552  & -1.50   & -0.88 & -0.960  & -1.343       \\
11& PD Benzene dimer              & 828  & -3.46   & -0.88 & -1.011  & -2.400       \\
12& Pyrazine dimer                & 736  & -5.13   & -2.40 & -2.428  & -3.897       \\
13& Uracil dimer                  & 920  & -11.21  & -8.46 & -7.478  & -9.306       \\
14& Stacked Indole-Benzene        & 989  & -5.65   & -2.13 & -2.096  & -4.244       \\
15& Stacked Adenine-Thymine       & 1127 & -13.61  & -9.23 & -8.431  & -11.128      \\
\hline                                     
  & Mixed                         &AOs  & THC-scGW & THC-scGWSOX & fcCCSD$^a$  & fcCCSD(T)$^b$    \\
\hline                                     
16& Ethene-Ethine                 & 322 & -1.67 & -1.47 & -1.244  & -1.431       \\
17& Benzene-H$_2$O                & 506 & -3.33 & -2.91 & -2.719  & -3.099       \\
18& Benzene-NH$_3$                & 529 & -2.38 & -1.82 & -1.793  & -2.183       \\
19& Benzene-HCN                   & 529 & -5.00 & -4.58 & -3.898  & -4.302       \\
20& T-shaped Benzene Dimer        & 828 & -3.60 & -2.12 & -1.856  & -2.552       \\
21& T-shaped Indole-Benzene       & 989 & -6.88 & -4.91 & -4.340  & -5.362       \\
22& Phenol Dimer                  & 920 & -7.96 & -6.44 & -5.823  & -6.706       \\
\hline
\hline
\end{tabular}

$^a$ Estimates computed in Ref.\cite{Sherrill:S22:2010} and retrieved from Biofragment Database.\cite{BFDb:2017}.

$^b$ Estimates computed in Ref.\cite{Sherrill:SCS-CCSD:NCI:2014} and retrieved from Biofragment Database.\cite{BFDb:2017}.
\end{table*}

All of the counterpoise-corrected interaction energies computed with scGF2 and scGF2dir are reported in Table~S2 in SI. Already for very small systems
scGF2 and scGF2dir significantly overestimate the interaction strength, sometimes by over 5 kcal/mol. 
This is consistent with OOMP2 results, 
although the extent of such an overestimation is notably more pronounced for scGF2.
A comparison with total energies in Table~\ref{tbl:tot_energies} indicates that this effect primarily originates from the dimer rather than the monomers.

The overestimation of the interaction energy is much more severe in larger systems (Table~S2 in SI). 
Interaction energies evaluated with scGF2dir are much larger than with scGF2, 
which is likely due to a lack of cancellation of Pauli exclusion principle violating terms resulting in a too large interaction energies. 
The overestimation of the interaction energies for both scGF2 and scGF2dir could be explained by lack of the renormalized or screened Coulomb interactions/integrals in the scGF2 and scGF2dir expressions.
Since the Coulomb interaction renormalization effectively reduces interactions, one can expect that it will reduce the intermolecular interaction in comparison with the bare perturbation theory. 
Recently, Furche and co-workers showed within a non-self-consistent framework that 
a bare perturbation theory with adiabatically connected symmetry adapted perturbation theory (AC-SAPT) 
can be diverging\cite{Furche:intermol:2020}, 
fully explaining numerous previous observations of deterioration of MP2 results for large systems, 
which is not apparent from MP2 calculations of small molecules. 
Our work shows that the full self-consistency uncovers this deterioration even for small molecules. 
We performed scGF2 and scGF2dir calculations only for a subset of molecules from S22 set because of convergence issues. 
For large molecules, scGF2dir leads to big changes in chemical potential leading to a slow logarithmic convergence of energy differences.

The renormalization of the interaction line by an infinite number of bubble insertions in the direct term (scGW) led to converging iterations in contrast to scGF2dir, which contains only one bubble insertion. 
In Table~\ref{tbl:int_energies}, we list interaction energies computed with scGW and scGWSOX and compare them 
with previously computed fcCCSD and fcCCSD(T) estimates. 
The errors of scGW are much smaller than of scGF2dir and scGF2 (see Tables~\ref{tbl:GW_NH3} and S2 in SI), 
which confirms the importance of including the renormalized interactions $W(\omega)$.
\begin{figure}[!h]
  \includegraphics[width=6cm]{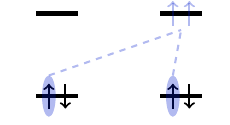} \hspace{2cm}
  \includegraphics[width=2cm]{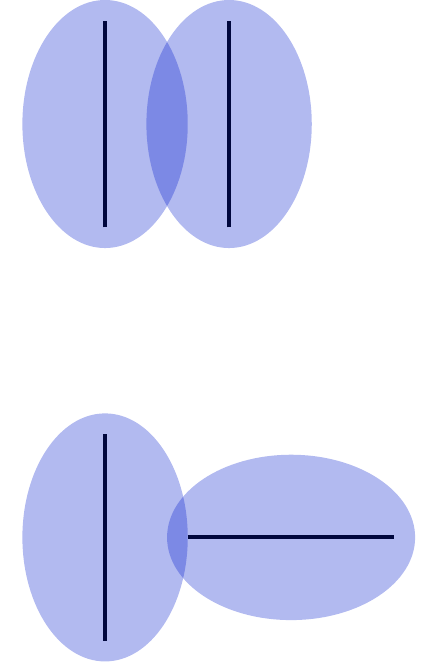}
\centering
\caption{Left: A correlated charge-transfer excitation violating the Pauli principle included in GW. Right: A schematic illustration of density overlaps in stacked and T-shaped dimers. 
The Pauli repulsion is more pronounced when more electron pairs can come into contact, leading to increased correlated Pauli repulsion for stacked dimers compared to T-shaped dimers.
         \protect\label{fig:EPV}
}
\end{figure}

Fully self-consistent scGW overestimates binding energies in all the cases except the smallest dimers: 
 (NH$_3$)$_2$, (H$_2$O)$_2$, and (CH$_4$)$_2$, for which the estimates are very close to CCSD(T).  
Such an overestimation is expected because of the violation of the Pauli exclusion principle due to a lack of exchange terms. Since the monomers are not very far from each other, there is a partial overlap of electron clouds making a partial charge transfer possible. We depict such contributions in Fig.\ref{fig:EPV}, left. Such same-spin charge-transfer contributions between monomers are not cancelled in scGW, which leads to an overestimation of the interaction. Additionally, for the same reason, the same-spin excitations within each monomer also can overestimate polarizability.

The correlated exchange term removes the exclusion principle violating same-spin contributions.
Our numerical results show that the inclusion of the bare second-order exchange term through scGWSOX reduces the strength of the intermolecular interactions in comparison to scGW 
for all systems from dispersion-dominated and mixed classes and some hydrogen-bonded systems. 
While scGW overbinds the S22 set, scGWSOX generally underbinds it with respect to CCSD(T) reference. 
This is expected because interactions in SOX are not renormalized resulting in an overall overcorrection. 
As shown in the previous sections, such an overcorrection raises the total energy resulting in  a more repulsive behavior for intermolecular interactions.
Due to short-ranged nature of exchange repulsive forces, their magnitude depends on the geometric orientation of the monomers, which is shown schematically in Fig.\ref{fig:EPV}, right. 
In the stacked orientation, the electron densities of the monomers overlap substantially, leading to both charge-transfer and Pauli repulsion forces. In the T-shaped orientation, the overlap of the densities is smaller causing a smaller Pauli repulsion.  
This idea is supported numerically as a difference between scGWSOX and scGW estimates for stacked and T-shaped benzene dimers and indole-benzene systems. The difference between scGW and scGWSOX estimates is always larger for the stacked orientations. Since SOX overestimates the correlated Pauli repulsion, the errors of interaction energy in the stacked orientation w.r.t CCSD(T) are larger than for the T-shaped orientation. We also observe the same trends for the stacked and H-bonded orientations of uracil dimers and adenine-thymine systems.
For all dispersion-dominated and mixed systems scGWSOX is closer to CCSD estimates than scGW, 
which is expected from the diagrammatic analysis of CCD. 
This is a promising result because it indicates that ladder insertions into the self-energy may not be significant for the description of dispersion forces.

The only systems, for which scGWSOX increased the strength of intermolecular interactions with respect to scGW, are water dimer, 
formic acid dimer, formamide dimer, and H-bonded uracil dimer. 
Although an exact origin of this increase is unclear, 
one can expect that the magnitude of the exchange repulsion would be smaller than for the dispersion-dominated systems 
because fewer electron pairs can come close enough to make Pauli repulsion as substantial as for the large flat stacked systems. 
Hydrogen bonds possess a complex nature. On one hand, they exhibit greater strength compared to typical dispersion interactions in small molecules. On the other hand, hydrogen bonds entail the redistribution of electron density. Consequently, an accurate depiction of hydrogen bonds necessitates the inclusion of higher-order diagrams, such as ladders, which describe electron correlation at a very short range.

\section{Conclusions and future work}
We have designed and implemented a novel MPI-parallel THC algorithm for the efficient evaluation of SOX. This advancement has enabled us to achieve full self-consistency in scGF2 and scGWSOX calculations with an order of magnitude larger number of orbitals than was previously feasible.
We have successfully applied the developed approach to evaluate intermolecular interactions using a supermolecular method. Our findings demonstrate that rigorous, fully self-consistent calculations are achievable without resorting to uncontrollable approximations.
Firstly, we achieved convergence of the interaction energies by varying the number of interpolation points in THC, demonstrating that $\alpha=10$ is adequate for resolving the interaction energies.
Secondly, we conducted an analysis of the total energies, comparing OOMP2 and scGF2 methodologies, and elucidated the behavior of the direct and exchange terms with both \emph{bare} and \emph{screened} interactions.
Finally, we meticulously benchmarked the implemented methods using the S22 set, ensuring thorough validation and accuracy assessment.
Our findings reveal that scGW typically overestimates the interaction strength, whereas scGWSOX tends to decrease it, particularly in cases where the number of contacting electron pairs of monomers is large.

The observed behavior contrasts with that of non-self-consistent RPA and SOSEX approximations, which tend to underbind the S22 set. Our comparison highlights significant variability of RPA concerning the chosen DFT reference. Although it is feasible to select a DFT functional for RPA that yields estimates closer to CCSD(T), such an empirical approach is not guaranteed to generalize to larger systems. In contrast to non-self-consistent methods, interaction strengths calculated using fully self-consistent methods possess a clear and well-defined physical interpretation.

We attribute the overestimation of interaction strength in scGW to the inclusion of exclusion-principle-violating diagrams. The SOX term, integrated within scGWSOX, exaggerates the magnitude of correlated Pauli repulsion, a trend particularly evident when examining stacked geometric orientations.

The scGWSOX estimates of intermolecular interactions closely align with CCSD estimates for dispersion-dominated and mixed systems, which is expected due to the shared terms between CCD and GW. This numerical similarity suggests the potential for future improvements within both Green's function and coupled-cluster theories.

In particular, the \emph{screened} interactions effectively reduce the interactions' strength, suggesting that incorporating screened interactions into the exchange correlated term by means of SOSEX or G3W2 could alleviate or potentially eliminate the overestimation of Pauli repulsion.
In the near future, we also plan to extend our analysis of intermolecular interactions and explore vertex insertions not only into the self-energy but also into the polarization bubble. 

\section*{Acknowledgments}
P.P. and D. Z. were supported by a grant from the Simons Foundation as part of the Simons Collaboration on the many-electron problem.
The computations in this work were run at facilities supported by the Scientific Computing Core at the Flatiron Institute, a division of the Simons Foundation. 

\section*{Supplementary Material}
MP2, scGF2, scGF2dir interactions energies of selected systems from S22 set.

\section*{Data Availability}

\section*{Appendix: Conditional size-extensivity of self-consistent Green's function methods}

Size-consistency and size-extensivity are important for evaluation of total energies and interaction energies. A size-consistent method gives total energies that are separable for non-interacting systems $E_{AB} = E_A + E_B$; thus, providing a correct behaviour in the non-interacting limit. A size-extensive method gives total energies that scale linearly with respect to $N$ identical non-interacting fragments $E_{supersystem} = N E_{fragment}$. 
 Although size-consistency is studied extensively for the wave-function methods\cite{Szabo_ostlund,BartlettShavitt:CC,OlsenText}, the discussion of Green's functions is often limited to size-extensivity of total energies and size-intensivity of excitation energies\cite{March:book:1967} only with rare exceptions\cite{Deleuze:size-cons:G:1995,Nooijen:size_cons:2005}. This is because traditionally Green's function methods are used for solids, for which only size-extensivity of methods can be defined in the reciprocal space. While one can even define size-consistency and size-extensivity directly in terms of Green's functions via linked-cluster theorem\cite{Deleuze:size-cons:G:1995}, such an approach may not be apparent for readers unfamiliar with Green's functions. The size-consistency of the Green's function approaches that we use can be non-obvious because of the reference-independent nature of fully self-consistent methods, fully removing the dependence on a reference determinant. 
 In this appendix, we provide a pedagogical discussion of size-consistency and size-extensivity for Green's function theories presented in this manuscript. We follow an algebraic style of derivation similar to the discussion of size-consistency for the wave-function methods\cite{OlsenText,Nooijen:size_cons:2005}.
It is instructive to consider two separate cases.

\textbf{Case 1: Fixed chemical potential}. 
Consider a supersystem consisting of $n$ non-interacting fragments, which are not necessarily identical. For such a supersystem, any one-electron or two-electron integrals are zero if at least one index belongs to a different fragment than other indices. Let us assume that we have a zero-temperature HF solution satisfying a multiplicative separability condition for the wave function $\ket{\Phi} = \ket{\Phi_1\cdots\Phi_n}$ in localized orbitals. 
Then the Green's function and the static self-energy matrix have a block-diagonal structure:
\begin{gather}
G^{HF} = 
\begin{pmatrix}
G_{\textbf{11}} & 0 & 0 & \dots \\
0 & G_{\textbf{22}} & 0 &  \dots \\
0 & 0 & G_{\textbf{33}} & \dots \\
\end{pmatrix} \qquad
\Sigma^{HF} = 
\begin{pmatrix}
\Sigma^{HF}_{\textbf{11}}[G^{HF}] & 0 & 0 & \dots \\
0 & \Sigma^{HF}_{\textbf{22}}[G^{HF}] & 0 &  \dots \\
0 & 0 & \Sigma^{HF}_{\textbf{33}}[G^{HF}] & \dots \\
\end{pmatrix},
\end{gather}
where $G_{\textbf{ii}}$ and $\Sigma^{HF}_{\textbf{ii}}$ are the Green's function and the static self-energy for each fragment. Since the energy expression contains only traces over the density matrices and self-energies, HF energy is additive for such solutions $E_{\infty}[G^{HF}] = \sum_{i=1}^n E_{\infty}[G^{HF}_{\mathbf{ii}}]$.

In our previous work, we found the  finite-temperature HF solutions that correspond to multiplicatively separable zero-temperature solutions. Such finite-temperature solutions preserve the separability of the Green's function and the static self-energy. Therefore, the total energy is also additive for them. The occurrence of these solutions can be seen from a convergent iterative process that we use in practice: starting from a separable zero-temperature static self-energy matrix, we find the right-hand-side of the Dyson equation, invert it, find the finite-temperature Green's function, construct the static self-energy matrix from it, etc. At all the iterative steps the separability of the self-energy and the Green's function is preserved because of the structure of the static self-energy expression and the Dyson equation. 

\begin{figure}[!h]
  \includegraphics[width=7cm]{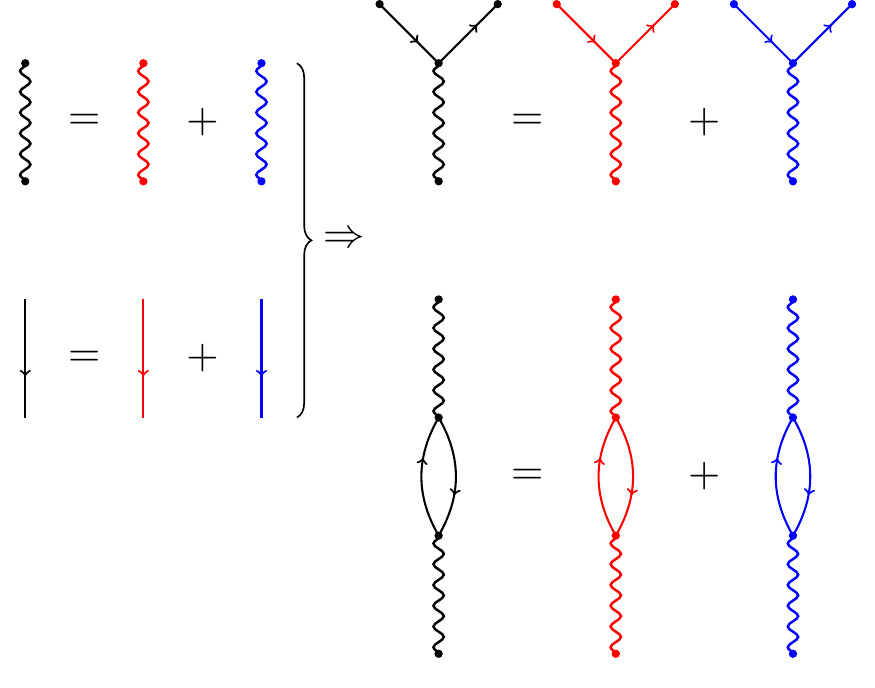}
\centering
\caption{Examples of separability of different quantities. Each quantity for the supersystem is shown in black color. Quantities for individual fragments are shown in other colors. Left: separability of the ERIs and the one-electron Green's function (not necessarily \emph{bold}). Right: immediate consequences of separability of ERIs and the Green's function. Any connected diagrams contracting integrals and the Green's function are separable. Some examples of such diagrams are shown here.
         \protect\label{fig:connected}
}
\end{figure}
Any self-energy skeleton diagram from \emph{bold} perturbative series has a connected structure. For example, all the diagrams entering GW and GWSOX self-energies are connected (Fig.~\ref{fig:diag}). Since the only non-zero ERIs are the ones with indices from the same fragment $(\textbf{ii}|\textbf{ii})$, any connected diagram with the separable Green's functions results in separable quantity\cite{note:separability}. Fig~\ref{fig:connected} shows examples of such contractions. For example, any diagram contributing to $W$ is connected; therefore, $W$ is separable as long as it is evaluated from the separable Green's function. The same holds for the self-energy diagrams that we consider
\begin{gather}
\Sigma^{dyn}[G] = 
\begin{pmatrix}
\Sigma^{dyn}_{\textbf{11}}[G] & 0 & 0 & \dots \\
0 & \Sigma^{dyn}_{\textbf{22}}[G] & 0 &  \dots \\
0 & 0 & \Sigma^{dyn}_{\textbf{33}}[G] & \dots \\
\end{pmatrix}
\end{gather}
Due to this separable property, once again the self-consistent iterations will provide separable self-energies and Green's functions at every iteration. The energy is additive due to the Galitskii--Migdal expression (Eq.~\ref{eq:E2b}). Therefore, for a fixed chemical potential finite-temperature fully self-consistent methods are size-consistent $E_{tot}[G] = \sum_i E_{tot}[G_{\textbf{ii}}]$. 

Physically, such an iterative process corresponds to an insertion of \emph{reducible} self-energies. Since such \emph{reducible} self-energies also have connected structure, they are separable and give separable Green's functions. 

\textbf{Case 2: Optimized chemical potential}. 
In practice, calculations performed without optimization of the chemical potential can result in a wrong average number of electrons. This is why we always find such a chemical potential $\mu$ that the average number of electrons remains unchanged. To do that we need to change the chemical potential during the iterative process. Starting from a separable self-energy, we can find separable Green's functions from the Dyson equation at \emph{any} value $\mu$. Therefore, we still get additive energies at every iteration with changing chemical potential of the supersystem. However, only the \emph{total} number of electrons in the supersystem is maintained. The local number of electrons for every fragment can change during iterations.

\textbf{Comment 1}. All the separability arguments above also apply for any linear combinations of separable self-energy or Green's function contributions. 
Therefore, we can safely use any algorithms relying on linear combinations of self-energies, such as damping and frequency-dependent CDIIS and LCIIS algorithms from Ref\cite{Pokhilko:algs:2022}. We would like to note that causality of the Green's functions and self-energies is not necessarily preserved by arbitrary linear combinations.

\textbf{Comment 2}. The iterative process is implicitly assumed to be stable, meaning that the solution does not fall to a different solution (that may not be separable) if numerical noise is present. 

\textbf{Comment 3}. The connected diagrammatic structure of self-energies for fully self-consistent methods also leads to a separability of the disconnected and connected (cumulant) components of the two-particle Green's functions and density matrices as clear from their derivation from Ref\cite{Pokhilko:tpdm:2021}. 
This fact also immediately leads to size-consistency of both one- and two-body energies. 

\textbf{Comment 4}. The initial separability condition of the HF solution is important. 
Considering stretched molecules with broken chemical bonds and starting from an RHF solution violating size consistency, 
we were able to converge restricted scGW and scGF2 solutions that also violate size consistency\cite{Pokhilko:local_correlators:2021,Pokhilko:algs:2022}. 
The reason why it happens lies in the charge-resonance contributions violating the multiplicative separability condition and leading to artificially large particle-number fluctuations. 
Such charge-resonance contributions are absent in the FCI limit.
While correlated fully self-consistent methods partly suppress the magnitude of such contributions\cite{Pokhilko:local_correlators:2021}, 
they are still there and are leading to significant errors. 
Considering instead UHF solutions maintaining the strong separability condition for a number of specific systems, 
 we were able to converge size-consistent unrestricted scGF2 and scGW solutions and even evaluate accurate effective magnetic couplings within the broken-symmetry approach~\cite{Pokhilko:local_correlators:2021,Pokhilko:BS-GW:solids:2022,Pokhilko:Neel_T:2022}. 
 More sophisticated cases may require
 finding a separable HF solution that breaks multiple symmetries\cite{Scuseria:GHF:2011}.

\renewcommand{\baselinestretch}{1.5}


\begin{thebibliography}{100}
\bibitem{note:GM}
The total energy can also be found from other expressions, which give the same
  energy value as the Galitskii--Migdal formula at the self-consistency.
\bibitem{note:crossing_sym}
In Green's function literature this permutation property is also known as
  ``crossing symmetry''.
\bibitem{note:ed}
In the physics community, the exact diagonalization (ED) is used to refer to
  the method known as the full configuration interaction (FCI) in the quantum
  chemistry community.
\bibitem{note:separability}
Although this statement is often considered as obvious in quantum chemistry, to
  ensure clarity, we repeat it here. Let us have a connected graph. Let us
  color its vertices and endges in a way that the each edge and the vertices
  that it connects have the same color. Then the color of all edges and all
  vertices in the graph are the same. The proof by contradiction is almost
  trivial. Suppose there are two vertices $A$ and $B$ that have different
  colors. Since the graph is connected, there is a path from the $A$ to $B$.
  Since all edges connected to the vertex must have the same color, we conclude
  that the path from $A$ to $B$ must have the same color as $A$. But since it
  is connected to $B$, it must also has a color of $B$. We arrive to
  contradiction. Therefore, all vertices have the same color. Since the edges
  have the same color as the vertices they connect, the whole graph has the
  same color.
\end{thebibliography}
\end{document}